\documentclass[%
 reprint,
superscriptaddress,
 amsmath,amssymb,
 aps,
]{revtex4-2}
\usepackage{graphicx}
\usepackage{dcolumn}
\usepackage{bm}
\usepackage[caption=false]{subfig}
\usepackage{amsmath}
\usepackage{comment}
\usepackage{calligra}
\DeclareMathAlphabet{\mathcalligra}{T1}{calligra}{m}{n}
\DeclareFontShape{T1}{calligra}{m}{n}{<->s*[2.2]callig15}{}

\usepackage{amssymb}
\usepackage{tabularx}
\usepackage{booktabs}
\usepackage{multirow}  
\usepackage{xcolor}
\usepackage{slashed}
\usepackage{url}
\usepackage{fontawesome5}
\usepackage{scalerel}
\usepackage{tikz}
\usetikzlibrary{svg.path} 
\definecolor{orcidlogocol}{HTML}{A6CE39}
\tikzset{
  orcidlogo/.pic={
    \fill[orcidlogocol] svg{M256,128c0,70.7-57.3,128-128,128C57.3,256,0,198.7,0,128C0,57.3,57.3,0,128,0C198.7,0,256,57.3,256,128z};
    \fill[white] svg{M86.3,186.2H70.9V79.1h15.4v48.4V186.2z}
                 svg{M108.9,79.1h41.6c39.6,0,57,28.3,57,53.6c0,27.5-21.5,53.6-56.8,53.6h-41.8V79.1z M124.3,172.4h24.5c34.9,0,42.9-26.5,42.9-39.7c0-21.5-13.7-39.7-43.7-39.7h-23.7V172.4z}
                 svg{M88.7,56.8c0,5.5-4.5,10.1-10.1,10.1c-5.6,0-10.1-4.6-10.1-10.1c0-5.6,4.5-10.1,10.1-10.1C84.2,46.7,88.7,51.3,88.7,56.8z};
  }
}

\newcommand\orcidicon[1]{\href{https://orcid.org/#1}{\mbox{\scalerel*{
\begin{tikzpicture}[yscale=-1,transform shape]
\pic{orcidlogo};
\end{tikzpicture}
}{|}}}}

\definecolor{mycolor}{RGB}{0,0,204}
\definecolor{pink}{RGB}{255,0,127}
\usepackage{hyperref}
\hypersetup{
  colorlinks   = true,    
  urlcolor     = pink,    
  linkcolor    =blue,    
  citecolor    = pink      
}
\usepackage{tabularx}
\usepackage{graphicx}
\usepackage{dcolumn}
\usepackage{bm}

\usepackage{float}
\begin{document}

\preprint{APS/123-QED}

\title{Exploring Primordial Black Holes and Gravitational Waves with R-Symmetric GUT Higgs Inflation} 

\author{Nadir Ijaz \orcidicon{0009-0007-2972-4511}}%
 \email{nadirijaz1919@gmail.com}%
\affiliation{Department of Physics, Quaid-i-Azam University, Islamabad 45320, Pakistan 
}%
\author{Mansoor Ur Rehman \orcidicon{0000-0002-1780-1571}}%
 \email{mansoor@qau.edu.pk}%
\affiliation{Department of Physics, Quaid-i-Azam University, Islamabad 45320, Pakistan 
}%
\date{\today}

\begin{abstract}
This study investigates the realization of R-symmetric Higgs inflation within the framework of no-scale-like supergravity, aiming to elucidate the formation of primordial black holes and observable gravitational waves within a class of GUT models. We explore the possibility of an ultra-slow-roll phase in a hybrid inflation framework, where the GUT Higgs field primarily takes on the role of the inflaton. The amplification of the scalar power spectrum gives rise to scalar-induced gravitational waves and the generation of primordial black holes. The predicted stochastic gravitational wave background falls within the sensitivity range of existing and upcoming gravitational wave detectors, while primordial black holes hold the potential to explain the abundance of dark matter.
Furthermore, we highlight the significance of the leading-order nonrenormalizable term in the superpotential of achieving inflationary observables consistent with the latest experimental data. Additionally, the predicted range of the tensor-to-scalar ratio, a key measure of primordial gravitational waves, lies within the observational window of future experiments searching for B-mode polarization patterns in cosmic microwave background data.

\end{abstract}

\maketitle


\section{\label{sec:level1}Introduction}
The detection of gravitational waves by experiments such as LIGO and Virgo~\cite{LIGOScientific:2016sjg,LIGOScientific:2017ycc,LIGOScientific:2017vox,LIGOScientific:2017bnn,LIGOScientific:2016aoc} has opened up new avenues for investigating primordial black holes (PBHs) \cite{Zeldovich:1967lct,Hawking1971,BJC1974,BJC1975,Khlopov1985}. 
Most recently, the 15-year pulsar timing data collected by the North American Nanohertz Observatory for Gravitational Waves (NANOGrav) presented convincing evidence for a low-frequency Gravitational Wave Background (GWB)~\cite{NANOGrav:2023gor}, potentially originating from PBHs~\cite{Afzal_2023}. Future space-based GW interferometers like LISA, BBO, and DECIGO~\cite{amaroseoane2017laser,Yagi:2011wg} are anticipated to detect similar background signals. 
PBHs have been proposed as potential constituents of dark matter, accounting for anywhere from a fraction to the entirety of its abundance. One of the primary mechanisms for PBH formation in the early universe involves the amplification of the primordial curvature perturbation spectrum. 
The mass of these PBHs is intricately linked to the Hubble mass, representing the mass energy contained within a Hubble sphere at the time of PBH formation. Consequently, PBHs offer a plausible explanation for a diverse range of cosmic phenomena \cite{Sasaki_2018, Carr_2020, Green_2021, escrivà2023primordial,_zsoy_2023, Villanueva_Domingo_2021}.

In the vast landscape of cosmology, inflation emerges as a crucial concept that sheds light on fundamental characteristics of the observable universe at large scales. Inflation explains critical aspects of the universe's structure, such as its immense size, uniformity, isotropy, and overall geometry. It gives rise to primordial density fluctuations, acting as the foundational ``seeds" for the creation of the cosmic structures we observe today, including galaxies and galaxy clusters. These primordial density fluctuations, in the standard inflationary scenario, originate from quantum fluctuations of a scalar field during the inflationary epoch.  Additionally, inflation triggers the amplification of tensor perturbations in the spacetime metric, resulting in the generation of primordial gravitational waves.

Given the effectiveness of cosmic inflation in elucidating the early universe, it's pertinent to consider inflation as a mechanism for magnifying the primordial power spectrum. Beyond the formation of primordial black holes, the amplified perturbations in scalar curvature are expected to yield significant secondary tensor perturbations. These aspects have been investigated in recent studies \cite{Di_2018,clesse2018detecting,Bartolo_2019,Hajkarim_2020,Liu_2020,Kawai_2023,Aldabergenov:2020bpt,Balaji:2023ehk,Ghoshal:2023wri,Ghoshal:2023pcx,Qin:2023lgo,Ijaz:2023cvc,Ballesteros:2022hjk,Braglia:2020taf,Braglia:2020eai,Braglia:2022phb}.

In the quest to establish a `standard model of inflationary cosmology,' supersymmetric hybrid inflation (SHI)~\cite{Dvali:1994ms,Copeland:1994vg,Linde:1997sj,Senoguz:2004vu,Buchmuller:2000zm,Rehman:2009nq} provides an elegant framework for integrating grand unified theories (GUTs) of particle physics \cite{Senoguz:2003zw}. Previous studies have explored the formation of PBHs within SHI, employing R-symmetry breaking terms in the K\"ahler potential \cite{Spanos:2021hpk,Ijaz:2023cvc}. Similar models have also been investigated in~\cite{Kawai_2023}.
This paper investigates the formation of PBHs in an R-symmetric SHI framework for a generic GUT model, without introducing any R-symmetry violating terms in either the superpotential or the K\"ahler potential. We adopt a no-scale-like supergravity-based framework, where the GUT Higgs field primarily functions as the inflaton. A multifield treatment is necessary for PBH formation. To achieve inflationary observables consistent with observational constraints, we also consider the leading-order nonrenormalizable term in the superpotential.

The paper is structured as follows: Section~\ref{Sec2} provides an overview of the R-Symmetric Higgs inflation within a generic GUT model, focusing on its formulation within a no-scale-like supergravity framework. In Section~\ref{Sec-III}, we delve into the effective single-field treatment of the model. Section~\ref{Sec4} extends the analysis to include the multifield treatment, exploring the presence of PBHs and gravitational waves, with particular attention given to the role of the leading-order nonrenormalizable term in the superpotential. We highlight the parametric range where observable inflationary primordial gravitational waves are expected, along with a brief discussion on successful reheating and leptogenesis scenarios. Additionally, we touch upon the potential of PBHs as dark matter and their relevance in explaining NanoGrav observations. Finally, Section~\ref{conclusion} presents our conclusions.

\section{\label{Sec2}R-Symmetric Higgs Inflation}

We consider an R-symmetric superpotential of the form,
\begin{equation}
    W = \kappa\, S\left(H^2-M^2\right) + \frac{\beta}{m_P} \,SH^3,
\end{equation}
where $\kappa$ and $\beta$ are dimensionless coupling, $M$ is some mass scale, and $m_P = 2.4\times10^{18}$~GeV is the reduced Planck mass. From here on, we will employ Planckian units, where the reduced Planck mass is taken as unity, $m_P=1$. The parameter $\beta$ can in general be complex but we assume it to be real and positive similar to $\kappa$ and $M$. The R-charge of the $H$-superfield is zero, and the R-charge of the $S$-superfield is identical to that of the superpotential $W$. 
The superpotential presented above has previously been utilized in shifted hybrid inflation models~\cite{Khalil:2010cp}, where the scalar component of the gauge singlet field $S$ serves as the inflaton, while the GUT Higgs field remains stabilized at a local minimum during inflation. However, in the current investigation, these roles will be reversed, with the GUT Higgs field primarily assuming the role of the inflaton.

The above form of superpotential can elegantly incorporate several GUT models listed below:
\begin{itemize}
    \item $SU(5)$ or $SU(5) \times U(1)$ where $H$ belongs to the adjoint representation $24$ ~\cite{Khalil:2010cp,Rehman:2018gnr,Pal:2019pqh,ahmed2023probing,Ijaz:2023cvc}.
    \item $SU(4)_c \times SU(2)_L\times U(1)_R$ where $H$ can be identified with the adjoint representation $15$~\cite{Afzal:2023cyp}.
\end{itemize}
While dealing with conjugate pair of GUT Higgs fields ($H \oplus \bar{H}$), the terms, $\kappa S H^2$ and $\beta \,S H^3$, in the above superpotential are replaced by $\kappa S H\bar{H}$ and $\beta \,S(H\bar{H})^2$:
\begin{itemize}
    \item $SO(10)$ with $16 \oplus \overline{16}$~\cite{Kyae:2005vg}.
    \item $SU(3)_c \times  SU(2)_L \times SU(2)_R \times SU(1)_{B-L}$ with $ (1, 1, 2, 1) \oplus (1, 1, 2, -1)$~\cite{Dvali:1997uq}
    \item Flipped $ SU(5) \equiv SU(5)\times U(1)$ with $(10,1) \oplus (\overline{10},-1)$~\cite{Rehman:2009yj,Civiletti:2013cra,Rehman:2018nsn,Rehman:2009nq,Rehman:2018gnr,Abid:2021jvn}.
    \item $SU(3)_c \times  SU(2)_L \times U(1)_Y \times U(1)_{B-L}$ with $ (1,1,0 , 1) \oplus (1,1,0,-1)$~\cite{Afzal:2022vjx,Rehman:2017gkm}. 
\end{itemize}

To realize non-minimal inflation we consider a  K\"ahler potential of the form $K=-3\ln\Phi$, where
\begin{equation}
\begin{split}
    \Phi&= 1-\frac{1}{3}(|S|^2+|H|^2)+\frac{\chi}{4}(H^2+h.c.)
    +\frac{\gamma_4}{3}|S|^4 \\& +\frac{\gamma_6}{3}|S|^6.
    \end{split}
\end{equation}
Here $\chi,\, \gamma_4 $ and $\gamma_6$ are dimensionless parameters. 
The quartic term is usually employed to stabilize the $S$ field at the origin during inflation \cite{Lee:2010hj, Senoguz:2004ky}. As discussed below, the sextic term plays an important role in generating primordial black holes. 
It's important to note that each term within the  K\"ahler potential preserves the R-symmetry, unlike models where primordial black holes are formed with a cubic term in the  K\"ahler potential, albeit at the cost of $R$-symmetry breaking at nonrenormalizable level \cite{Ijaz:2023cvc}.

The scalar-gravity portion of the Lagrangian is expressed in the following manner
\begin{equation}
    \mathcal{L}_J = \sqrt{-g_J}\left[\frac{\Phi\mathcal{R}_J}{2}-g_J^{\mu\nu} \mathcal{G}_{ij} \partial_{\mu}z^i\partial_{\nu}z^{*j}-V_J \right],
\end{equation}
where $g_J^{\mu\nu}$  is the inverse of the Jordan frame spacetime metric $g^J_{\mu\nu}$, $\mathcal{R}_J$ is the Ricci scalar in the Jordan frame,  $V_J$ represents the scalar potential in the Jordan frame, and  $z_i\in \{H,\,S\}$.  The metric on the field-space manifold in the Jordan frame, $\mathcal{G}_{ij}$, is defined as 
\begin{equation}
\mathcal{G}_{ij}\equiv-3\frac{\partial^2\Phi}{\partial_{z_i}\partial_{z_j^*}}.
\end{equation}
The fields $S$ and $H$ can be written in the form,
\begin{equation}
      S=\frac{1}{\sqrt{2}}s\, e^{i\theta_s}, \qquad H=\frac{1}{\sqrt{2}}h\, e^{i\theta_h},
\end{equation}
where $s$ and $h$ are the real canonically normalized scalar fields with respective phases  $\theta_s$ and phase $\theta_h$.
With these definitions, the above Lagrangian takes the following form,
\begin{eqnarray} \label{e5}
\mathcal{L}_J &=& \sqrt{-g_J} \left[\frac{\Phi\mathcal{R}_J}{2}-\frac{1}{2}g_J^{\mu\nu}(\partial_{\mu}h\partial_{\nu}h +  h^2 \partial_{\mu}\theta_h \partial_{\nu}\theta_h)  \right. \nonumber \\
&-&  \left.  \frac{\zeta}{2}g_J^{\mu\nu} (\partial_{\mu}s\partial_{\nu}s + s^2 \partial_{\mu}\theta_s \partial_{\nu}\theta_s)-V_J \right],   
\end{eqnarray} 
where
\begin{align}
\zeta&=1-2\gamma_4s^2-\frac{9}{4}\gamma_6\,s^4,
\end{align}
and
\begin{align}
    \Phi&=1-\frac{s^2}{6}+\frac{\gamma_4}{12}s^4+\frac{\gamma_6}{24}s^6
    +  \left( \frac{\chi}{4} \cos(2\theta_h) -\frac{1}{6} \right) \,h^2.
\end{align}

\begin{figure}[t!]\centering
\includegraphics[width=0.48\textwidth]{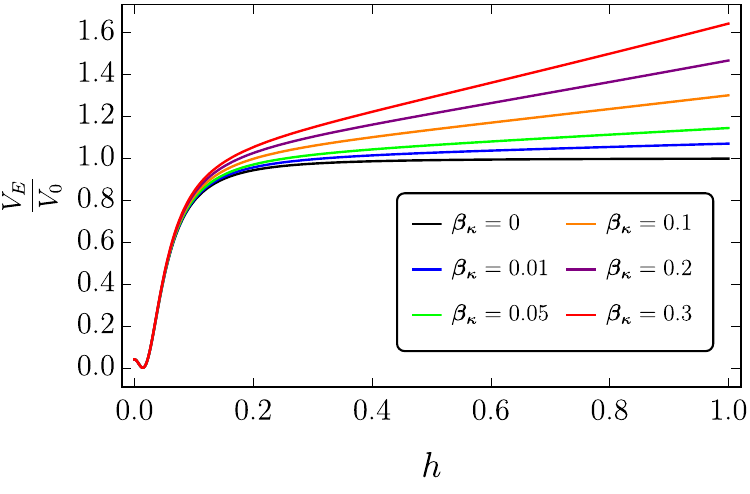}
\caption{\label{fig:betavsh} The normalized Einstein frame potential, with $V_0=\frac{\kappa^2}{4\,\xi^2}$, at $s=0$ for $M=0.01$, $\xi=10^3$, $\gamma_4=0.01$, $\gamma_6=0$ and different values of $\beta_{\kappa}$, where $\beta_{\kappa}=\beta/\kappa$.}
\end{figure}

We then proceed with a conformal transformation of the spacetime metric, to work in the Einstein frame
\begin{eqnarray}
    g^E_{\mu\nu} = \Phi \, g^J_{\mu\nu}.
\end{eqnarray}
This transformation leads to the following Lagrangian in the Einstein frame with a canonical coupling to gravity:
\begin{equation}
    \mathcal{L}_E=\sqrt{-g_E}\left[\frac{1}{2}\mathcal{R}_E-\frac{1}{2}G_{ij}g^{\mu\nu}_E \partial_{\mu}z^i\partial_{\nu}z^{*j} - V_E\right],
\end{equation}
where in the Einstein frame, the kinetic term for the scalar fields incorporates a field-space metric \cite{Kaiser_2010,Kaiser_2013},
\begin{equation}
    G_{ij}=\frac{1}{\Phi}\left[\mathcal{G}_{ij}+\frac{3}{2} \frac{\Phi_{,i}\Phi_{,j}}{\Phi}\right],
\end{equation}
where $\Phi_{,i}\equiv\partial \Phi/\partial z^i$.
Along the D-flat direction, the supergravity (SUGRA)  scalar potential in the Einstein frame, $V_E=\Phi^{-2}V_J$, is defined as,
\begin{equation}
V_E = e^{K/m_P^2}\left((K^{-1})_{ij} D_{z_i}WD_{z^*_j}W^*-3m_P^{-2}\left|W\right|^2\right),
\end{equation}
where,
\begin{equation}
    D_{z_i}W\equiv\frac{\partial W}{\partial z_i}+m_P^{-2}\frac{\partial K}{\partial z_i},\quad K_{ij}\equiv\frac{\partial^2K}{\partial z_i\partial z^*_j},
\end{equation}
$D_{z^*_i}W^*=(D_{z_i}W)^*$ and $K=-3\ln\Phi$. For the background equations and the relevant power spectra in a multifield framework, see Refs. \cite{Kaiser_2013, Peterson_2011, Gordon_2000, Nakamura_1996, Gong_2011,Geller:2022nkr}.

A discussion is in order regarding the stabilization of $s$, $\theta_s$ and $\theta_h$.  For simplicity, we initially disregard terms involving the $\beta$ and $\gamma_6$ couplings. In the limit where $s \gg 1$ and $\chi h^2 \gg 1$, the Einstein-frame scalar potential simplifies to:
\begin{eqnarray}
V_E &\simeq&  \frac{4\,\kappa^2}{\chi^2 \cos^2(2\theta_h)} \left[ 1 - \frac{8}{\chi 
 h^2 \cos(2\theta_h)} + 2 \gamma_4 s^2 \right.  \nonumber \\
 &+& \frac{16}{3} \frac{s^2}{\chi h^2} \left( \frac{1}{\cos(2\theta_h)} - \cos(2\theta_h)    \right) \nonumber \\
 &+&  \frac{4}{3} \frac{s^2}{\chi h^2} \left. \left( - \cos(2\theta_s) + 3 \tan(2\theta_h) \sin(2\theta_s) \right)  \right].     
\end{eqnarray}
This potential is minimized when $s=0$, $\theta_s = 0$ and $\theta_h = 0$.
To assess the stability of this multifield framework we calculate the following mass matrix~\cite{Ellis:2019bmm} :
\begin{equation}
M_{\text{scalar}}^2 =  \begin{bmatrix}
(K^{-1})_{ki} \mathcal{D}_{z_k^*}\partial_{z_j} V_E & (K^{-1})_{ki} \mathcal{D}_{z_k^*}\partial_{z_j^*} V_E \\
(K^{-1})_{ik} \mathcal{D}_{z_k}\partial_{z_j} V_E & (K^{-1})_{ik} \mathcal{D}_{z_k}\partial_{z_j^*} V_E
\end{bmatrix},
\end{equation}
where $\mathcal{D}$ represents the covariant derivative defined in~\cite{Ellis:2019bmm}. Upon diagonalizing this matrix, we obtain the following scalar masses for the real and imaginary components:
\begin{equation}
\begin{aligned}
M_{\text{Re}[S]}^2 &\simeq (3\gamma_4 \chi h^2 -2) \mathcal{H}^2, \\
M_{\text{Im}[S]}^2 &\simeq (3\gamma_4 \chi h^2 +2) \mathcal{H}^2, \\
M_{\text{Re}[H]}^2  &\simeq -\frac{24}{\chi h^2} \mathcal{H}^2, \\
M_{\text{Im}[H]}^2 &\simeq 4 \mathcal{H}^2,
\end{aligned}
\end{equation}
where $\mathcal{H}^2 \simeq 4\kappa^2 / (3 \chi^2) $ is the Hubble mass-squared. The condition $M_{\text{Re}[S]}^2 > 0$ is satisfied for $\gamma_4 > \frac{2}{3\chi h^2}$, ensuring that the field $s$ acquire a mass of order Hubble scale during inflation,  allowing it to quickly stabilize at the origin. Similarly,  $M_{\text{Im}[S]}^2 > 0 $ implies that the phase $\theta_s$ can also be stabilized at zero within a Hubble time.
The negative value $M_{\text{Re}[H]}^2<0$ is consistent with slow-roll inflation driven by $h$, resulting in a spectral index $n_s<1$. Meanwhile, the phase $\theta_h$, being associated with the heavy mas $M_{\text{Im}[H]}^2$, settles quickly to zero.

Introducing a nonzero $\beta$ value adds only minor corrections to the potential, so the stabilization remains effective, as confirmed by numerical checks. This stabilization also holds in the presence of a nonzero $\gamma_6$ term.
These results are utilized in the following sections.

\section{\label{Sec-III}Effective Single Field Inflation }
The field $S$ can be stabilized at zero which becomes a minimum with a vanishing $\gamma_6$ and a suitable nonzero value of $\gamma_4$. We thus obtain an effective Higgs potential which, for field values below $m_P$, can be employed to realize nonminimal Higgs inflation. This effective potential has been previously analyzed in great detail in \cite{Abid:2021jvn,Ijaz:2023cvc}. Also, see \cite{Masoud:2019cen} for a generalized treatment with $Z_n$ symmetry. In the present work, we aim to study the impact of leading order non-renormalizable term with parameter $\beta$ in the superpotential on inflationary observables. 

\begin{table}
\begin{tabularx}{\columnwidth}{|>{\centering\arraybackslash}X|>{\centering\arraybackslash}X|>{\centering\arraybackslash}X|>{\centering\arraybackslash}X|}
    \hline
    \multicolumn{4}{|c|}{For $N_0=50$} \\
    \hline
    \hline
    $\beta_{\kappa}$ & $\kappa$ & $n_s$  & $r$ \\
    \hline
    \hline
    $0$ & $0.0053$ & $0.961$ & $0.0043$ \\
    \hline
    $0.01$ & $0.0055$ & $0.964$ & $0.0046$ \\
    \hline
    $0.02$ & $0.0057$ & $0.966$ & $0.0051$ \\
    \hline
    $0.05$ & $0.0065$ & $0.974$ & $0.0066$ \\
    \hline
    \multicolumn{4}{|c|}{For $N_0=60$} \\
    \hline
    $0$ & $0.0044$ & $0.968$ & $0.0030$ \\
    \hline
    $0.01$ & $0.0047$ & $0.971$ & $0.0034$ \\
    \hline
    $0.02$ & $0.0049$ & $0.973$ & $0.0038$ \\
    \hline
    $0.03$ & $0.0052$ & $0.976$ & $0.0042$ \\
    \hline
\end{tabularx}
\caption{{\label{Tab1}}The predicted values of inflationary parameters with gauge symmetry breaking scale $M = 0.01$, $\xi=10^3$, $\gamma_4=0.01$ and $\gamma_6 = 0$.}
\end{table}

With $s$ stabilized at the origin, the scalar potential in the Einstein frame takes the following form
\begin{equation}
    V_E=\frac{1}{16}\frac{\left(2\,\kappa\,h^2-4 \, \kappa \,M^2+\sqrt{2}\,\beta \, h^3\right)^2 }{\left(1+\xi h^2\right)^2},
\end{equation}
with
\begin{equation}
\xi\equiv\frac{\chi}{4}-\frac{1}{6}.
\end{equation}
The effect of $\beta$ on the scalar potential is shown in Fig.~\ref{fig:betavsh} for fixed values of $\gamma_4 = 0.01$, $\xi = 10^3$ and $ M = 0.01$. 
An increase in the potential's slope is observed which is expected to introduce intriguing effects on the predictions of inflationary observables, as elaborated below.

It is important to mention that the $R$-symmetry is broken in the hidden sector once superpotential acquires a vev. 
This breaking is assumed to be communicated through gravity mediation, manifesting in the potential as soft SUSY breaking terms associated with the gravitino mass, $m_{3/2}$, which we have assumed to be around a few TeV. We have chosen to omit these terms since their impact is negligible during inflation. For instance, by comparing the soft mass term $m_{3/2}^2 h^2$ with the quartic term $\kappa^2 h^4$,  it becomes clear that the soft mass term can be neglected, as $h \gg m_{3/2}/ \kappa$ during inflation.

After conformal rescaling the canonically normalized inflaton field $\hat{h}(h)$ in the Einstein frame becomes a function of field $h$ as
\begin{equation}
    J(h)\equiv\left(\frac{d\hat{h}}{dh}\right)=\sqrt{\frac{1}{\Phi}+\frac{3}{2}\left(\frac{d\ln\Phi}{dh}\right)}.
\end{equation}
The slow-roll parameters can now be expressed in terms of $h$ as
\begin{equation}
    \epsilon(h)=\frac{1}{2}\left(\frac{V_E^{\prime}}{JV_E}\right)^2,\quad \eta(h)=\left(\frac{V_E^{\prime\prime}}{J^2V_E}-\frac{J^{\prime}V_E^{\prime}}{J^3V_E}\right),
\end{equation}
where a prime denotes a derivative with respect to $h$. The scalar spectral index $n_s$ and the tensor to scalar ratio $r$ to the first order in slow-roll approximation are given by

\begin{figure}[t!]\centering
\subfloat[\label{3DP}]{\includegraphics[width=0.45\textwidth]{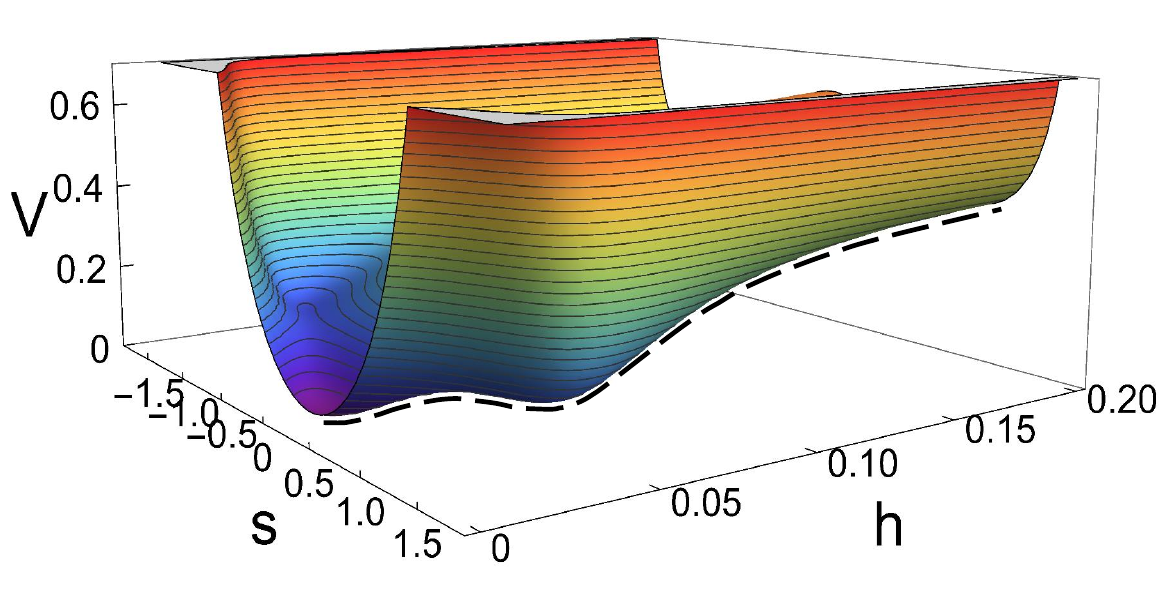}}\qquad
\subfloat[\label{3DT}]{\includegraphics[width=0.45\textwidth]{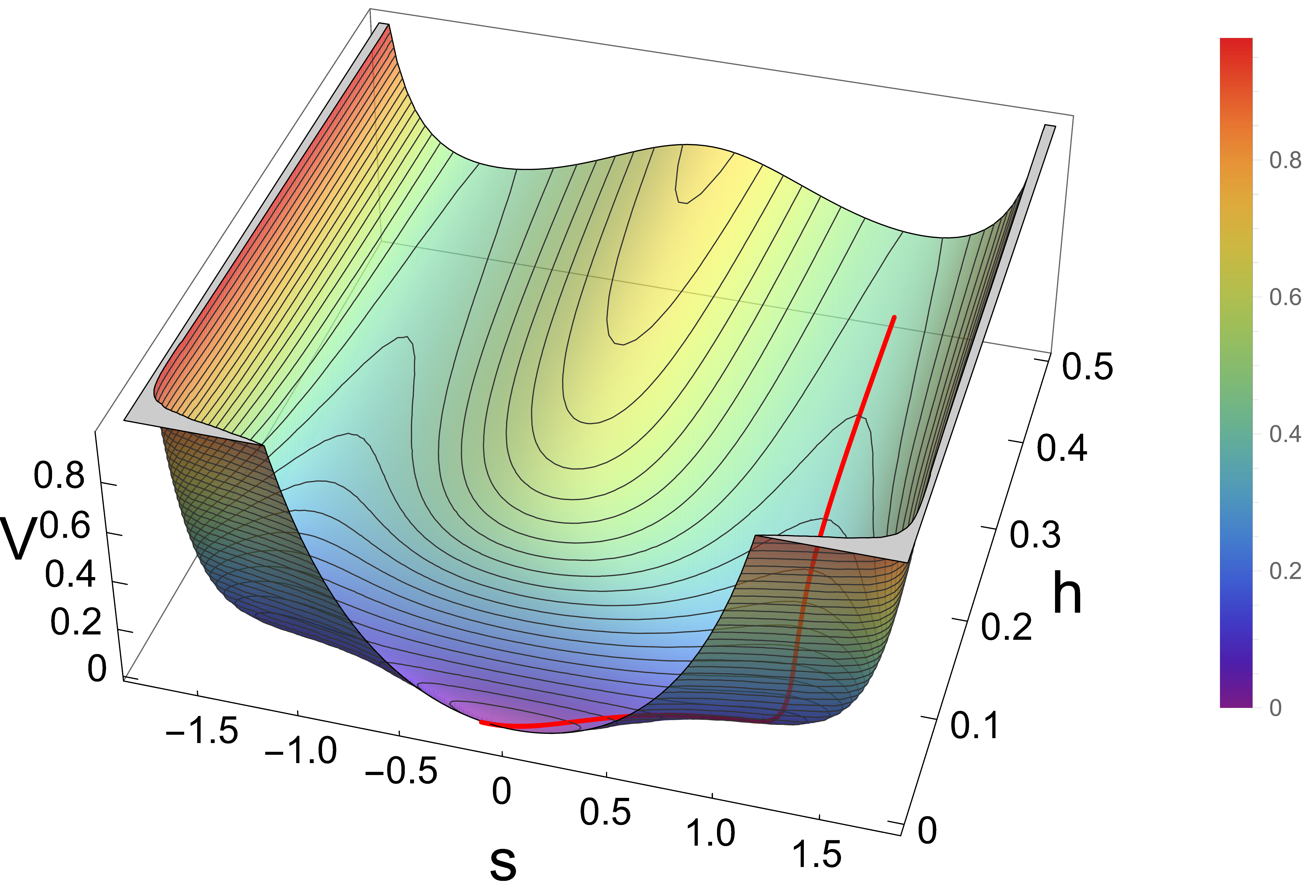}}
\caption{\label{3D} (a) The inflection point is depicted in the 3D Einstein frame potential. (b) The inflationary trajectory (red) superimposed on the Einstein frame potential. }
\end{figure}

\begin{equation}
    n_s\simeq1-6\epsilon(h_0)+2\eta(h_0),\quad r\simeq 16\epsilon(h_0),
\end{equation}
where the field value, $h_0$, corresponds to the number of e-folds,
\begin{equation}
    N_0=\frac{1}{\sqrt{2}}\int_{h_e}^{h_0}\frac{J(h)}{\sqrt{\epsilon(h)}}dh,
\end{equation}
before the end of inflation at $h=h_e$ defined by the condition $\epsilon(h_e)=1$. Also, $h_0$ corresponds to the pivot scale where the amplitude of the scalar power spectrum is normalized by Planck \cite{planck2020} to be,
\begin{equation}\label{eq20}
   A_s(k_0)=\left.\frac{V_E(h)}{24\pi\epsilon(h)}\right|_{h(k_0)=h_0}=2.137\times10^{-9},
\end{equation}
at $k_0=0.05~\text{Mpc}^{-1}$. 

Using eqs. (15), and (17), the slow-roll parameters in the large $\xi$ limit and leading order in $\beta$ are given by
\begin{equation}
\begin{aligned}
\epsilon&\simeq \frac{4}{3\psi^4}+\frac{2\beta_{\kappa}}{3\psi}\sqrt{\frac{2}{\xi}},\\
\eta&\simeq -\frac{4}{3\psi^2}+\frac{\beta_{\kappa}}{3}\sqrt{\frac{1}{2\xi}}\psi.
\end{aligned}
\end{equation}
Here, $\psi\equiv\sqrt{\xi}h$ is a dimensionless field variable taking the value at the pivot scale approximated as $\psi_0 \simeq \sqrt{4N_0/3}$ and at the end of inflation as $\psi_e \simeq (4/3)^{1/4} < \psi_0$. The spectral index and tensor-to-scalar ratio are then expressed as,
\begin{equation}
\begin{aligned}
n_s&\simeq 1-\frac{8}{3\psi_0^2}+\frac{2\beta_{\kappa}}{3}\frac{\psi_0}{\sqrt{2\xi}}\simeq 1-\frac{2}{N_0}+\frac{2\beta_{\kappa}}{3}\sqrt{\frac{2N_0}{3\xi}},\\
r&\simeq\frac{12}{N_0^2}+16\beta_{\kappa}\sqrt{\frac{2}{3N_0\xi}}.
\end{aligned}
\end{equation}
These analytical expressions elucidate how the scalar spectral index $(n_s)$ and the tensor-to-scalar ratio $r$ rise with increasing $\beta$. The influence of the parameter $\beta$ is detailed in Table~\ref{Tab1} for $50$ and $60$ e-folds, with a fixed value of $M = 0.01$. This role of $\beta$ also proves instrumental in attaining observationally consistent values of the scalar spectral index within the multifield scenario, as elaborated in the subsequent section.

\begin{figure}[t!]\centering
\includegraphics[width=0.48\textwidth,height=0.8\textwidth]{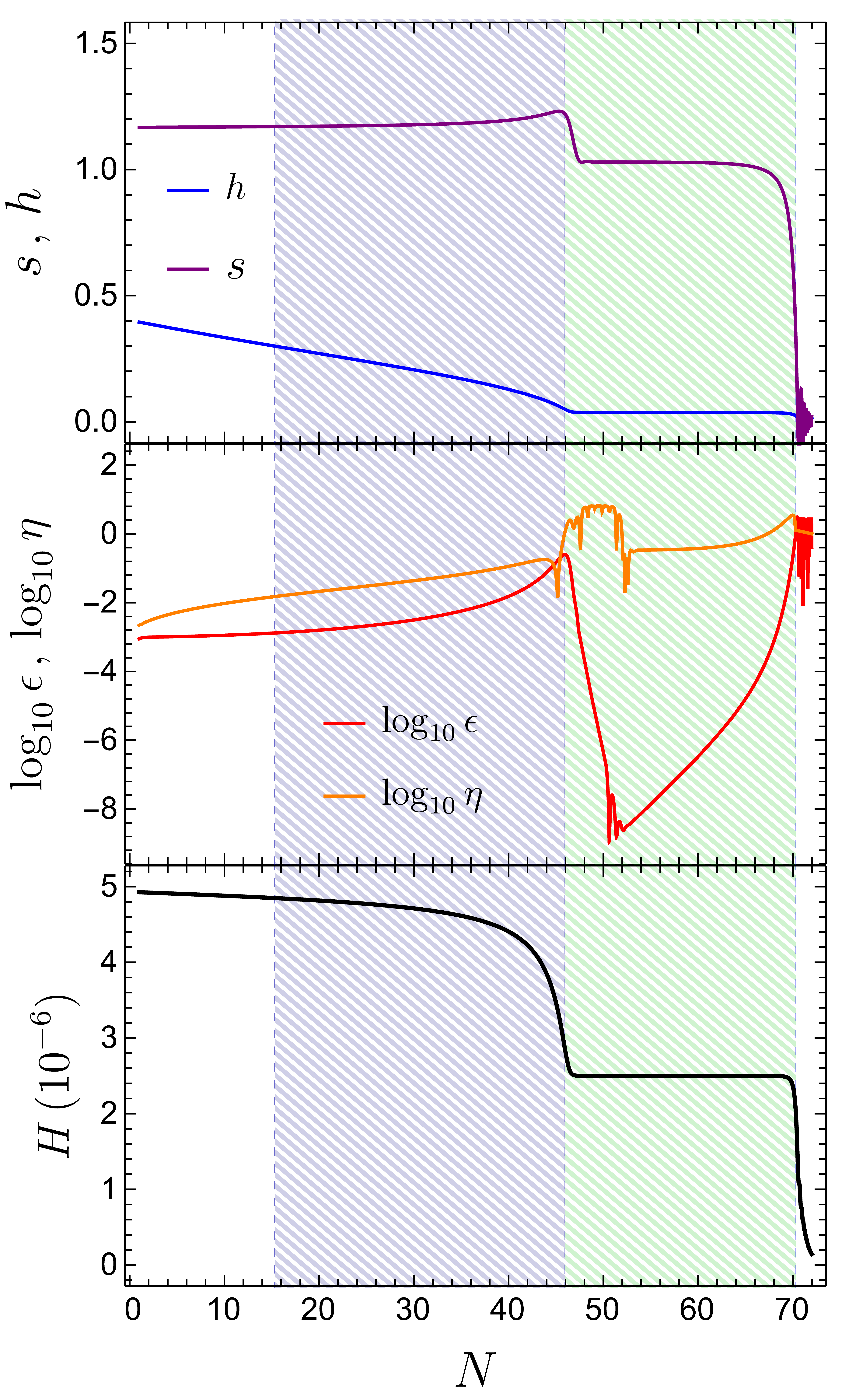}
\caption{\label{fig:FS} From top to bottom we show as a function of the number of e-folds, the dynamics of the fields, the slow-roll parameters ($\epsilon$ and $\eta$), and the Hubble parameter $H$. The blue-hatched region represents the first stage of inflation, and the green-hatched region represents the second stage.}
\end{figure}

\section{\label{Sec4}Multifield Inflation}
Setting $\gamma_6$ to zero enabled us to achieve effective single-field inflation, as investigated in the preceding section. However, in this section, we introduce a non-zero $\gamma_6$. For $\gamma_6>0$ and $\gamma_4<0$, two additional local minima emerge at non-zero values of the $s$-field for $\xi \gg 1$ and $h \gg M$, given approximately by:
\begin{equation}
s_{\text{min}} \simeq  \pm \frac{2}{3} \sqrt{-\frac{\gamma_4}{\gamma_6}}.
\end{equation}
With $\gamma_4<0$, the $s=0$ trajectory transforms into a local maximum for $h \gg M$. The significance of $\gamma_4$ is evident from the expression above, as it can shift the positioning of the local minimum (valley) of the potential towards larger $s$-field values at $h \gg M$ for a fixed $\gamma_6$. By selecting suitable values for $\gamma_6$ and $\gamma_4$, a region of false vacuum can be established near an effective inflection point in the potential~\cite{Kawai_2023}. This is the region where a phase of Ultra Slow Roll (USR) inflation can be realized. In Fig.~\ref{3DP}, we illustrate a 3D representation of the Einstein frame potential, showcasing the presence of this inflection point region. By appropriately tuning $\gamma_6$ and $\gamma_4$, the depth of this false vacuum can be modified. For our numerical computations, we utilized the multifield formalism tools as described in Refs.\cite{Kaiser_2013, Peterson_2011, Gordon_2000, Nakamura_1996, Gong_2011,Geller:2022nkr}.

In Fig.~\ref{3DT}, we present the inflationary trajectory superimposed on the Einstein frame potential, where $\gamma_6=0.13$ and $\gamma_4=-0.395804$. 
The role of the $s$-field requires some clarification, which can be achieved by examining the inflationary trajectory. 
Initially, when $h$ is large, the $s$-field forms a valley, and as the inflaton moves along this valley during this ``slow-roll'' phase, the $s$-field remains fixed at a particular value. However, at a certain point, the $h$-field encounters a sharp turn while the $s$-field is still fixed; this marks the beginning of the $s$-field's dynamic role. The $s$-field then proceeds to roll down a path characterized by an inflection point, with the $h$-field remaining nearly constant. This path ultimately leads to the minimum of the potential at $s=0$ and $h\sim M$. Thus, the formation of the inflection point is not due to a single field alone but rather a combination of both fields, with the $s$-field playing the dominant role.

The evolution of the fields is also depicted in the top panel of Fig.~\ref{fig:FS}. This trajectory undergoes two distinct stages: initially, it experiences typical slow-roll inflation within the valley, and upon reaching the inflection point, a second stage of Ultra Slow Roll (USR) inflation begins. During USR, the first slow-roll parameter, $\epsilon=-\dot{H}/H^2$, is significantly suppressed ($\epsilon_{USR}\ll\epsilon_{SR}$).
We identify the end of the first stage from the local maximum of $\epsilon$ (or when $\eta\equiv 2\epsilon-\dot{\epsilon}/2H\epsilon$ first crosses unity). Inflation ends when the parameter $\epsilon$ attains unity, as depicted in the middle panel of Fig.~\ref{fig:FS}. The blue-hatched region signifies the first stage of inflation, while the green-hatched region represents the second stage. The duration of USR can be regulated by adjusting the parameter $\gamma_4$ while keeping $\gamma_6$ fixed. The Hubble parameter is presented in the bottom panel of Fig.\ref{fig:FS}.

\subsection{The Role of $\beta$ Parameter}

The USR phase induces an enhancement of the scalar power spectrum. However, when $\beta=0$, prolonging the USR phase causes a shift in the scalar spectral index towards lower values and the tensor-to-scalar ratio towards larger values, moving away from the central range preferred by the Planck data. Table~\ref{Tab2} demonstrates how, for a fixed $\gamma_6$, the parameter $\gamma_4$ governs the duration of the USR phase and its corresponding predictions for inflationary observables.

\begin{table}
    \begin{tabular}{|>{\centering\arraybackslash}p{1cm}|>{\centering\arraybackslash}p{1.2cm}|>{\centering\arraybackslash}p{2cm}|>{\centering\arraybackslash}p{1.2cm}|>{\centering\arraybackslash}p{1cm}|>{\centering\arraybackslash}p{1.3cm}|}
        \hline
          & $\gamma_6$ & $\gamma_4$ & $\Delta N_2$ & $n_s$  & $r$ \\
    \hline
    \hline
    P-1 & $0.13$ & $-0.394700$ & $12$ & $0.956$ & $0.0059$\\
    \hline
     P-2 & $0.13$ & $-0.394830$ & $21$ & $0.944$ & $0.0095$\\
    \hline
    P-3 &  $0.13$ & $-0.394835$ & $26$ & $0.935$ & $0.0126$ \\
    \hline
    \end{tabular}
    \caption{{\label{Tab2}} Points for $\beta=0$, a fixed $\gamma_6$ and three different values of $\gamma_4$, showing how the parameter $\gamma_4$ controls the duration of the USR phase ($\Delta N_2$). The spectral index in this is inconsistent with the observational data and decreases with the increase in the duration of the USR phase. The scalar power spectrum for these points is shown in Fig.~\ref{fig:PS00}}
\end{table}

\begin{figure}[t!]\centering
\includegraphics[width=0.48\textwidth]{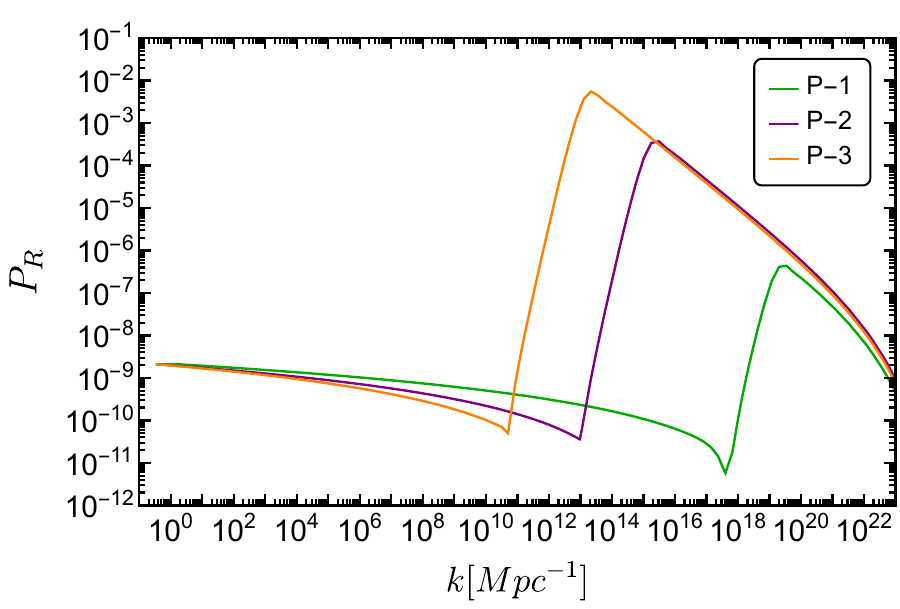}
\caption{\label{fig:PS00} Power spectrum of scalar perturbations, $P_R$, as a function of the comoving wavenumber $k$ computed numerically for points in Table~\ref{Tab2}}
\end{figure}

In our numerical analysis in this section, we set the reheating temperature at $10^{9}$ GeV, resulting in approximately $55$ e-folds of inflation. The power spectrum of curvature perturbations is computed numerically using a publicly available Mathematica package~\cite{Dias:2015rca} based on the transport method~\cite{Mulryne_2010, Mulryne_2011}. Fig.~\ref{fig:PS00} illustrates the power spectrum for the points listed in Table~\ref{Tab2}. We utilize the parameter $\kappa$ to normalize the power spectrum to the observed value (Eq.~\ref{eq20}) by Planck \cite{planck2020}. It's worth noting that for the case where $\beta=0$, although an enhancement in the power spectrum can be achieved, the scalar spectral index deviates from the observationally preferred 1-$\sigma$ region.

As demonstrated in Section~\ref{Sec-III}, the parameter $\beta$ exerts a significant effect by shifting the spectral index towards a more favored region. This prompts us to explore its potential to attain observationally consistent results for inflationary observables. Thus, we examine four Benchmark Points (BP) listed in Table~\ref{Tab3}, each featuring non-zero values of $\beta$, and present the resulting power spectra in Fig.~\ref{fig:PSF}. We incorporate constraints on the power spectrum from Planck~\cite{planck2020}, Lyman-alpha~\cite{Lyman}, and FIRAS $\mu$-distortions measurements~\cite{Fixsen_1996}. For a comprehensive investigation into spectral distortions in both single and multifield inflation scenarios, see Ref.~\cite{Baur:2023naq}. The constraints from spectral distortions on the scalar power spectrum are discussed in detail in~\cite{Tagliazucchi:2023dai}.

To quantify the fine-tuning required for the peak amplitude of the power spectrum, we adopt the method introduced in~\cite{Azhar:2018lzd} and further applied in~\cite{Cole:2023wyx}. In~\cite{Cole:2023wyx} the authors conduct a comprehensive analysis of the parametric dependence in various single-field inflation models, assessing the level of tuning needed for specific observables to remain within defined ranges. This measure is expressed as the logarithmic derivative of a given observable $\mathcal{O}$ with respect to the logarithm of any parameter $p$:

\begin{equation}\label{Eqq24}
    \epsilon_{\mathcal{O}}=\frac{d \log \mathcal{O}}{d \log p }
\end{equation}

\noindent Using Eq.~\ref{Eqq24} we calculated the fine tuning ($\epsilon_{peak}$) given in Table~\ref{Tab222}. The fine tuning required is of similar order as reported in~\cite{ Cole:2023wyx, Garcia-Bellido:2017mdw, Hertzberg:2017dkh}.

As illustrated in Table~\ref{Tab3}, leveraging non-zero values of $\beta$ allows us to readily attain scalar spectral index values within the 1-$\sigma$ bound of Planck data. This underscores the significance of the tensor-to-scalar ratio achieved within this model, a point further elaborated upon in the subsequent subsection.

\subsection{Primordial Gravitational Waves}
An important contributor to the stochastic gravitational wave background arises from the quantum fluctuations of the spacetime metric itself. While these tensor perturbations undergo amplification during inflation, direct detection remains challenging; nevertheless, they leave distinct imprints on the CMB spectrum via B-mode polarization. The tensor-to-scalar ratio, $r$, serves as a fundamental metric for these primordial gravitational waves, constrained from above by current CMB observations to $ r \leq 0.036$ at a 95\% confidence level~\cite{BICEP:2021xfz}. Future experiments are poised to explore the observable range,  $0.001 \lesssim r \lesssim 0.04$~\cite{EUCLID:2011zbd,SimonsObservatory:2018koc,LiteBIRD:2020khw}. Depending on the value of $\beta$, our model predicts the tensor-to-scalar ratio in the range, $0.02 \lesssim r \lesssim 0.04$. This range is consistent with the 1-$\sigma$ bound on $n_s$ and remains detectable by future experiments. Note that the effective single-field treatment,  where primordial black hole production is not considered, yields a lower value for $r$, potentially as low as $0.003$, a typical prediction of the Starobinski inflationary model.
The Benchmark Point (BP-4) listed in Table~\ref{Tab3} predicts a value of $r\simeq 0.1$, which falls beyond the 2-$\sigma$ bound of Planck data. Nonetheless, its significance will be addressed in the final subsection.

\subsection{Reheating and Leptogenesis}
Following the end of inflation, the inflaton begins oscillating around the global minimum defined by:
\begin{equation}
\langle S \rangle = 0, \, \langle H^2-M^2+\beta_{\kappa} H^3 \rangle = 0,
\end{equation}
where $\langle H \rangle \approx M$ within the relevant parametric range. The oscillating inflaton is composed of two complex scalar fields $S$ and $H$, both possessing a common mass. During this phase, the inflaton can decay into the lightest right-handed neutrino, $N$, through
\begin{equation}
W \supset \frac{\gamma_{ij}}{m_P}H^2N_iN_j \supset M^R_{ij} N_iN_j,
\end{equation}
where $ M^R_{ij} \sim \gamma_{ij} \langle H^2 \rangle / m_P$ constitutes the mass matrix for right-handed neutrinos, and $N$ represents the field with the lightest mass eigenvalue. Note that this term respects R-symmetry, with $N_i$ carrying an R-charge of one. After the reheating process, the lepton asymmetry generated by inflaton decay undergoes partial conversion to baryon asymmetry via sphaleron processes~\cite{Khlebnikov:1988sr,Harvey:1990qw}. To circumvent the issue of gravitino overproduction, as previously discussed, the reheating temperature is fixed at $10^9$~GeV. With this chosen reheating temperature, the observed baryon asymmetry can also be accounted for \cite{Ijaz:2023cvc}.

\begin{table}
    \begin{tabular}{|>{\centering\arraybackslash}p{1cm}|>{\centering\arraybackslash}p{1.2cm}|>{\centering\arraybackslash}p{2cm}|>{\centering\arraybackslash}p{1cm}|>{\centering\arraybackslash}p{1cm}|>{\centering\arraybackslash}p{1.4cm}|}
        \hline
          & $\gamma_6$ & $\gamma_4$ & $\beta_{\kappa}$ & $n_s$  & $r$ \\
    \hline
    \hline
     BP-1 &  $0.13$ & $-0.395804$ & $0.12$ & $0.965$ & $0.021$ \\
    \hline
     BP-2 & $0.09$ & $-0.301128$ & $0.16$ & $0.964$ & $0.028$\\
    \hline
     BP-3 & $0.065$ & $-0.235523$ & $0.22$ & $0.964$ & $0.040$\\
    \hline
    BP-4 & $0.04$ & $-0.1631133$  & $0.65$ &  $0.963$ & $0.100$\\
    \hline
    \end{tabular}
    \caption{{\label{Tab3}} The Benchmark Points (BP) we used for our numerical analysis, along with the predictions for the inflationary observable $n_s$ and $r$. Here the first benchmark point has the same $\gamma_6$ as in table~\ref{Tab2}.}
\end{table}

\begin{table}

    \begin{tabular}{|>{\centering\arraybackslash}p{2cm}|>{\centering\arraybackslash}p{2cm}|>{\centering\arraybackslash}p{2cm}|>{\centering\arraybackslash}p{2cm}|}
        \hline
         & $\gamma_6$  & $\gamma_4$ & $\beta$ \\[1mm]
        \hline
        \hline
        Fiducial & $0.13$  & $-0.395804$ & $0.12$ \\
        \hline
        $\epsilon_{P_{\text{Peak}}}$ & $-1.4\times 10^6$  & $-7.8\times10^5$& $-6.8\times 10^3$ \\[1mm]
        \hline
    \end{tabular}
    \caption{{\label{Tab222}} Fiducial and fine-tuning parameters for the model.}
\end{table}

\begin{figure}[t!]\centering
\includegraphics[width=0.48\textwidth]{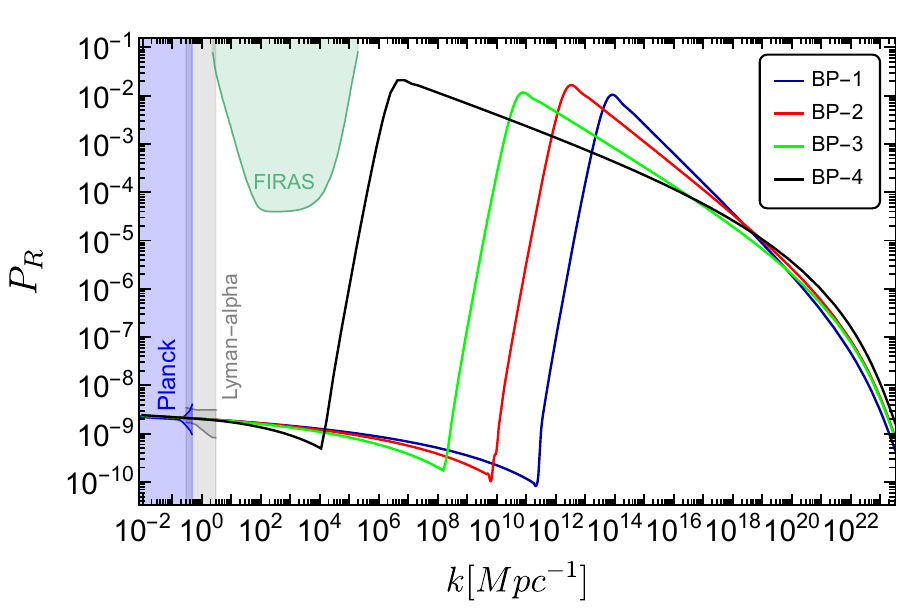}
\caption{\label{fig:PSF} Power spectrum of scalar perturbations, $P_R$, as a function of the comoving wavenumber $k$ computed numerically for benchmark points in Table~\ref{Tab3}. The shaded regions are  excluded by Planck~\cite{planck2020}, Lyman-alpha~\cite{Lyman}, and FIRAS $\mu$-distortions measurements~\cite{Fixsen_1996}. }

\end{figure}

\subsection{Abundance of Primordial Black Holes}
During horizon reentry, the enhanced curvature perturbations can cause the formation of PBHs through gravitational collapse. Our focus in this section is to compute the mass of PBHs and their fractional energy density abundances, assuming they are formed during the radiation-dominated epoch.

When the perturbation associated with the PBH enters the horizon, the PBH's mass is dictated by the horizon's mass. The relationship between the scale of the perturbation and the mass of the PBH during formation is expressed as follows,
\begin{equation}
M_{PBH}=\gamma M_{H,0}\Omega_{rad,0}^{1/2}\left(\frac{g_{*,0}}{g_{*,f}}\right)^{1/6}\left(\frac{k_0}{k_f}\right)^2,
\end{equation}
where $M_{H,0}=\frac{4\pi}{H}$ represents the mass of horizon, $\Omega_{rad}$ is used to denote the energy density parameter for radiation, while $g_*$ represents the effective degrees of freedom. The subscripts $f$ and $0$ correspond to the time of formation and the present day respectively. The ratio between the PBH mass and the horizon mass is denoted by $\gamma$, which is estimated to be  $\gamma\simeq3^{-3/2}$, in the simple analytical result \cite{carr1975primordial}.
The energy density of the PBHs at present can be determined by redshifting it at the formation time. This means that $\rho_\text{PBH,0}=\rho_\text{PBH,f}(a_\text{f}/a_0)^3\approx \gamma \beta \rho_\text{rad,f}(a_\text{f}/a_0)^3$, as the PBHs act like matter.  Here $\beta$ denotes the mass fraction of the Universe collapsing in PBH mass.

\begin{figure}[t!]\centering
\includegraphics[width=0.48\textwidth,height=0.35\textwidth]{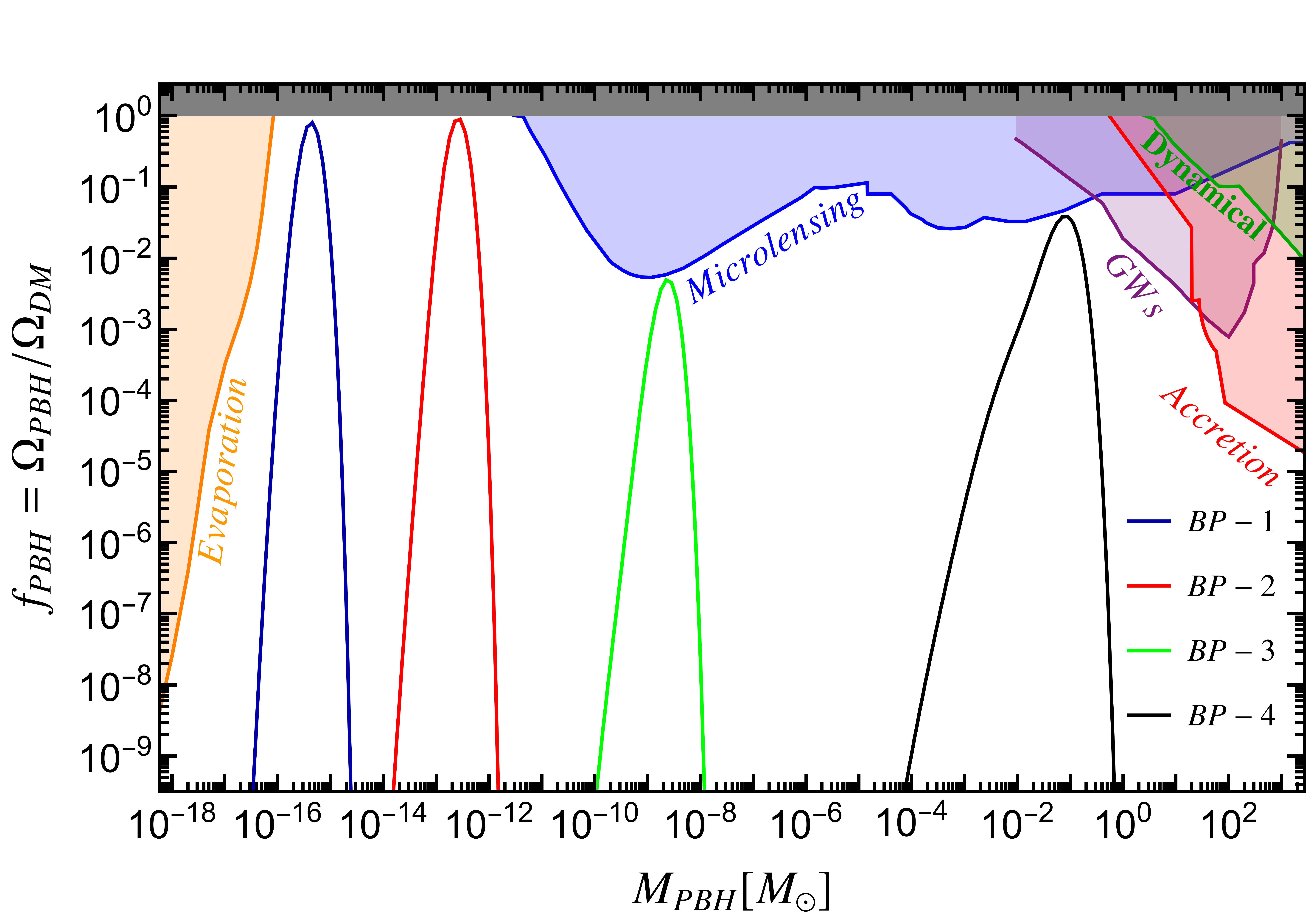}
\caption{\label{fig:PBHDM} Fraction of dark matter in the form of PBHs with mass $M_{PBH}$.
For the various constraints see Ref. \cite{Green_2021} 
and for an updated repository of data see  \href{https://github.com/bradkav/PBHbounds}{{\color{pink}\faGithub}/bradkav/PBHbounds}. }
\end{figure}

The Press-Schechter method is used to evaluate the mass fraction $\beta$ assuming that the overdensity $\delta$ follows a gaussian probability distribution function. The collapse of PBH is determined by a threshold value denoted as $\delta_c$. Thus, the mass fraction can be expressed as

\begin{equation}
    \beta=\int_{\delta_c}^{\infty}d\delta\frac{1}{\sqrt{2\pi\sigma^2}}\exp\left(-\frac{\delta^2}{2\sigma^2}\right),
\end{equation}
where $\sigma$ is the variance of curvature perturbation that is related to the co-moving wavenumber \cite{Young_2014}.
\begin{equation}
    \sigma^2=\frac{16}{81}\int^{\infty}_{0}d\ln q (q k^{-1})^4W^2(q k^{-1}))\mathcal{P}_{\zeta}(q),
\end{equation}
where $W^2(q k^{-1})$ is a window function that we approximate with a Gaussian distribution function,
\begin{equation}
    W(q k^{-1})=\exp\left[-\frac{1}{2}(q k^{-1})^2\right].
\end{equation}
For $\delta_c$ we assume values in the range between $0.4$ and $0.6$ \cite{Harada_2013,Musco_2009,Musco_2005,Escriv__2020,Escriv__2021,Musco_2021}.

The total abundance is $\Omega_\text{PBH,tot}=\int d\ln M_{PBH} \Omega_\text{PBH}$, with $\Omega_\text{PBH}$ expressed in terms of $f_{\text{PBH}}$ which is given by,
\begin{align}
    f_{\text{PBH}}&\equiv\frac{\Omega_{\text{PBH,0}}}{\Omega_{\text{CDM,0}}}\nonumber\\
    &\approx\left(\frac{\beta(M)}{8.0\times10^{-15}}\right)\left(\frac{0.12}{\Omega_{\text{CDM,0}}h^2}\right)\left(\frac{\gamma}{0.2}\right)^{3/2}\nonumber\\&
    \times \left(\frac{106.75}{g_{*,f}}\right)^{1/4} \left(\frac{M_{PBH}}{10^{20} \text{g}}\right)^{-\frac{1}{2}},
\end{align}
where  $\Omega_{CDM,0}$ is the today's density parameter of the cold dark matter and $h$ is the rescaled Hubble rate today.
The fractional abundance of the primordial black holes as a function of their mass is shown in Fig.~\ref{fig:PBHDM}. We have used the benchmark point parameters of table~\ref{Tab3}. For only the first two BPs, the PBHs can explain dark matter in totality. The shaded regions correspond to the various observational constraints \cite{Carr_2010, Carr_2021,Green_2021}.

\begin{figure}[t!]\centering
\includegraphics[width=0.5\textwidth, height=0.35\textwidth]{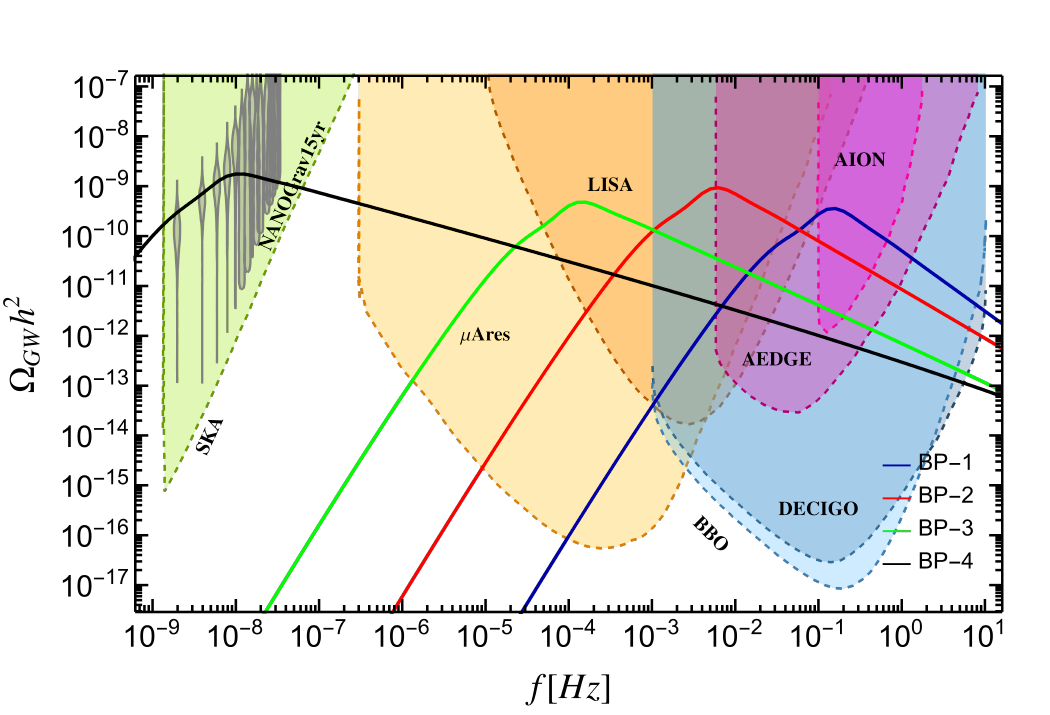}
\caption{\label{fig:GWS} Fraction of the energy density in GWs relative to the critical energy density of the universe as a function of the frequency. We show the sensitivity curves of: AION \cite{Badurina_2020}, AEDGE \cite{El_Neaj_2020}, DECIGO \cite{Kudoh_2006, kawamura2020current}, BBO \cite{Harry_2006}, LISA \cite{amaroseoane2017laser}, $\mu-$Ares \cite{Sesana_2021}, SKA \cite{ska1, ska2} and the gray violins are the NANOGrav frequency bins \cite{Afzal_2023}. We use as a comparison the first 15 bins of the NANOGrav 15 signal.}
\end{figure}

\subsection{Scalar-induced gravitational wave}
The GWs energy density within the subhorizon regions can be expressed as \cite{Maggiore_2000},
\begin{equation}\label{4.12}
    \rho_{\text{GW}}=\frac{\langle \overline{\partial_lh_{ij}\partial^lh^{ij}}\rangle}{16a^2},
\end{equation}
here the average over oscillation is denoted by the overline and $h_{ij}$ is the tensor mode. We can decompose $h_{ij}$ as,  
\begin{equation}\label{4.13}
    h_{ij}(t,\textbf{x})=\int \frac{d^3\textbf{k}}{(2\pi)^{3/2}}\left( h_\textbf{k}^{+}(t)e_{ij}^{+}(\textbf{k})+h_{\textbf{k}}^{\times}(t)e_{ij}^{\times}(\textbf{k})\right)e^{i\textbf{k}.{x}},
\end{equation}
with $e_{ij}^{+}(\textbf{k})$ and $e_{ij}^{\times}(\textbf{k})$ being the polarization tensors and can be expressed as,  
\begin{align}
 e_{ij}^{+}(\textbf{k})&=\frac{1}{\sqrt{2}}\left(e_{i}(\textbf{k})e_{j}(\textbf{k})-\bar{e}_i(\textbf{k})\Bar{e}_j(\textbf{k})\right),\\
  e_{ij}^{\times}(\textbf{k})&=\frac{1}{\sqrt{2}}\left(e_{i}(\textbf{k})\Bar{e}_{j}(\textbf{k})-\bar{e}_i(\textbf{k})e_j(\textbf{k})\right),
\end{align}
here $e_{i}(\textbf{k})$ and $\Bar{e}_{j}(\textbf{k})$ are orthongal to each other. Expression (\ref{4.13}) can be used in (\ref{4.12}), which yields, 
\begin{equation}
    \rho_{\text{GW}}(t)=\int d\ln k\frac{\overline{\mathcal{P}_h{(t,k)}}}{2}\left(\frac{k}{2a}\right)^2.
\end{equation}
In this expression $\mathcal{P}_{h}\equiv\mathcal{P}_h^{+,\times}$ and can be defined as, 
\begin{align}
    \langle h_{\textbf{k}}^{+}(t)h_{\textbf{q}}^{+}(t)\rangle&=\frac{2\pi^2}{k^3}\mathcal{P}_{h}^{+}(t,k)\delta^3(\textbf{k}+\textbf{q}),\\
     \langle h_{\textbf{k}}^{\times}(t)h_{\textbf{q}}^{\times}(t)\rangle&=\frac{2\pi^2}{k^3}\mathcal{P}_{h}^{\times}(t,k)\delta^3(\textbf{k}+\textbf{q}).
\end{align}

It should be noted that $\mathcal{P}_h^{+}(t,k)=\mathcal{P}_h^{\times}(t,k)$. The energy density parameter for gravitational wave is defined as follows,
\begin{equation}
    \Omega_{\text{GW}}(t,k)\equiv\frac{\rho_{\text{GW}}(t,k)}{\rho_{\text{crit}}}=\frac{1}{24}\left(\frac{k}{aH}\right)^2\overline{\mathcal{P}_h(t,k)},
\end{equation}
where we have omitted the polarization index. 

The following tensor perturbation equation should be solved for obtaining the tensor power spectrum  \cite{Baumann_2007,Kohri_2018,Inomata_2017},
\begin{equation}
    h^{\prime\prime}_{\textbf{k}}+2aHh_{\textbf{k}}^{\prime}+k^2h_{\textbf{k}}=4S_{\textbf{k},}
\end{equation}
where derivatives with respect to conformal time are denoted by the prime. The source term $S_k$  is expressed as,
\begin{equation}
    \begin{split}
        S_k=\int&\frac{d^3q}{\sqrt[3]{(2\pi)}}q_iq_je_{ij}(\textbf{k})\left.\biggr[2\Phi_{\textbf{q}}\Phi_{\textbf{k}-\textbf{q}} \right. \\ & \left.
        +\frac{4}{3+3\omega}\left(\frac{\Phi^{\prime}_{\textbf{q}}}{\mathcal{H}}+\Phi_{\textbf{q}}\right)\left(\frac{\Phi^{\prime}_{\textbf{k}-\textbf{q}}}{\mathcal{H}}+\Phi_{\textbf{k}-\textbf{q}}\right) \right],
    \end{split}
\end{equation}
the $\omega$ here is the equation of state. The generation of induced gravitational waves is assumed to occur during the radiation-dominated era. Thus, at the time of their generation, we have \cite{Kohri_2018},
\begin{equation}
\small
    \begin{split}
        \Omega_{\text{GW}}(t_f,k) &= \frac{1}{12} \int_0^{\infty} dv \int_{|1-v|}^{1+v} du \left( \frac{4v^2-(1+v^2-u^2)^2}{4uv}\right)^2 \\ & \times \mathcal{P}_{\zeta}(ku) \mathcal{P}_{\zeta}(kv) \left(\frac{3(u^2+v^2-3)}{4u^3v^3} \right)^2 \\ &
        \times \left.\biggr[\left(\pi^2(-3+v^2+u^2)^2 \theta(-\sqrt{3}+u+v) \right) \right. \\ & \left.+\left(-4uv+(v^2+u^2-3)\text{Log} \left| \frac{3-(u+v)^2}{3-(u-v)^2} \right| \right)^2 \right].
    \end{split}
\end{equation}
We can determine the energy density parameter for the current time as follows \cite{Kohri_2018, Ando_2018},

\begin{equation}
    \Omega_{\text{GW}}=\Omega_{rad,0}\Omega_{\text{GW}}(t_f)
\end{equation}
where we have multiplied $\Omega_{\text{GW}}(t_f)$ with the today radiation energy density parameter, $\Omega_{rad,0}$. 

Utilizing the numerically computed power spectrum of scalar perturbations, as depicted in Fig.~\ref{fig:PSF}, we derive the corresponding second-order gravitational wave (GW) spectrum. Our findings, illustrated in Fig.~\ref{fig:GWS}, represent the energy density fraction of gravitational waves relative to the critical energy density, denoted as $\Omega_{GW}$. Our analysis reveals that the predictions for all four benchmark points (BPs) intersect the sensitivity thresholds of upcoming experiments. Notably, the prediction for the fourth BP (depicted by the black curve) is also consistent with the NanoGrav 15-year data. However, it's worth noting that the tensor-to-scalar ratio for this BP contradicts the observations from the Planck satellite. Consequently, the constraints imposed by the observational data on the tensor-to-scalar ratio hinder our model's ability to fully account for the NanoGrav observations.

\section{Conclusion}  \label{conclusion}
We have successfully implemented GUT Higgs inflation within an R-symmetric hybrid framework based on no-scale supergravity. In the single-field treatment, our inflationary predictions for the scalar spectral index and the tensor-to-scalar ratio align closely with those of the Starobinski model, albeit with some variation. However, the multifield scenario presents a distinct prediction of observable primordial gravitational waves, with a larger tensor-to-scalar ratio falling within the detectable range of future experiments.
The inclusion of the leading-order non-renormalizable term in the superpotential is crucial for ensuring the consistency of multifield predictions with Planck data. Additionally, our framework easily accommodates successful reheating and leptogenesis. An ultra-slow-roll phase is attainable within the multifield framework, leading to an enhancement of the scalar power spectrum. This enhancement facilitates the production of primordial black holes, offering a potential explanation for the abundance of dark matter.
Lastly, the amplified scalar power spectrum gives rise to scalar-induced gravitational waves, which are detectable by both existing and forthcoming gravitational wave detectors.
\\  

\nocite{*}

\bibliography{NMHI}
\end{document}